# Atomic Resonant Tunneling in the Surface Diffusion of H Atoms on Pt(111)

Cheng Bi[1,2] and Yong Yang[1,2*]

1. *Key Lab of Photovoltaic and Energy Conservation Materials, Institute of Solid State Physics, HFIPS, Chinese Academy of Sciences, Hefei 230031, China.*
2. *Science Island Branch of Graduate School, University of Science and Technology of China, Hefei 230026, China.*

**ABSTRACT:**

The quantum motions of hydrogen (H) atoms play an important role in the dynamical properties and functionalities of condensed phase materials as well as biological systems. In this work, based on the transfer matrix method and first-principles calculations, we study the dynamics of H atoms on Pt(111) surface and numerically calculate the quantum probability of H transferring across the surface potential fields. Atomic resonant tunneling (ART) is demonstrated along a number of diffusion pathways. Owing to resonant tunneling, anomalous rate of transfer is predicted for H diffusion along certain path at low temperatures. The role of nuclear quantum effects (NQEs) on the surface reactions involving H is investigated, by analyzing the probabilities of barrier-crossing. The effective barrier is significantly reduced due to quantum tunneling, and decreases monotonically with temperature within a certain region. For barrier-crossing processes where the Van't Hoff-Arrhenius type relation applies, we show the existence of a nonzero low-temperature limit of rate constant, which indicates nontrivial activities of H-involved reactions at cryogenic conditions.

-------------------------------------------------------------------------------------------------------------------
*Corresponding Author: Y. Yang (yyanglab@issp.ac.cn).

## 1. Introduction

For quantum many-body systems like polyatomic molecules and condensed matter, the Schrödinger equation which governs the motions of the microscopic particles, can only be solved through various approximations instead of obtaining the exact solutions. The most commonly used method is the adiabatic approximation, also known as the Born-Oppenheimer approximation [1]. This procedure decouples the kinetic degrees of freedom between electrons and ions. In practice, most of the electronic structure calculations solve the Schrödinger equation of electrons in a given ionic potential field self-consistently by using numerical methods in which the atomic nuclei or ions are viewed as classical particles. The approximation greatly reduces the computational cost, but it inevitably ignores the nuclear quantum effects (NQEs) such as nuclear zero-point energy, tunneling, and coherence [2-6], especially when dealing with systems containing light-weight atoms.

The earliest studied NQEs in the condensed matter systems may be the quantized collective excitations of lattice vibrations in crystals, i.e., phonons [7]. In fact, these studies can be traced back to the pioneering works of Einstein and Debye on the specific heat of solids under low temperature conditions [8]. In addition to the elementary excitations of the collective motions of lattice atoms, another aspect of NQEs is their spatial delocalization as single microscopic particles due to the intrinsic nature of wave-particle duality. In this case, to properly describe the interactions and properties of a many-particle system, not only the wave functions of the electrons but also those of the nuclei/ion cores are required. For example, the NQEs in hydrogen bonding systems which are known as proton-sharing, result from the overlap of proton wave functions. Early researches of NQEs based on Feynman's path integral formulation of quantum mechanics involve the simulations of light atoms in bulk state such as hydrogen in metals [9] and quantum matter like condensed helium [10]. In recent decades, along with the development of advanced theoretical and simulation techniques (such as path integral molecular dynamics, PIMD [11, 12]) and experimental techniques (such as deep inelastic neutron scattering [13-15], inelastic electron tunneling spectroscopy [16-19]), the researches on hydrogen-rich or hydrogen bonding systems such as liquid hydrogen, water, and some biological molecules [20-23] have made the exploration of NQEs gradually reenter the mainstream.

Notably, current research works mainly focus on NQEs related to the vibrational properties of atoms and molecules [18], the real space bonding structures and momentum distribution of hydrogen in bulk condensed phases (such as aqueous solution environment) [24-26], and the transport of protons [27-29]. Studies of NQEs on the surface diffusion and transport of atoms are still very limited, although such processes are crucial, especially for the surface reactions involving hydrogen [30]. On the other hand, the *ab initio* PIMD (APIMD) whose interatomic potentials are constructed using *ab initio* methods [31], despite its reliability and accuracy, has to consume huge computing resources because of the need to evaluate all possible paths. Therefore, APIMD is usually restricted to simulations of small systems consisting of a dozen atoms or less. Atomic simulations using APIMD on surfaces have been reported [32, 33]; nevertheless, APIMD studies of the surface diffusion of H on surfaces are still scarce due to the huge computational burden involved. Although simulations using APIMD for systems containing several tens to even more than one hundred atoms are possible [34], it is still too demanding since massively parallel supercomputing facilities are required [34]. Alternatively, approximated methods based on Feynman path integral formalism, such as ring polymer molecular dynamics (RPMD) using model potentials [35-37], or other approximated methods [38] were employed to study the diffusion of H atoms on metal surfaces [39]. In such studies, only single barriers are considered; studies on the more realistic situation which includes multiple barriers (e.g., double barriers) or surface potential wells are still lacked. For double barriers, a remarkable quantum effect is the resonant tunneling of particles, in which the microscopic particles pass through the energy barriers with a probability of 100% when their kinetic energies are below the barrier height. The phenomenon of resonant tunneling of atoms, or *atomic resonant tunneling* (ART), has *not* been reported in the studies using APIMD, RPMD or other approximated methods based on path integral. This may be due to the intrinsic deficiency of these methods, or practical difficulties in explicitly including the quantum effects of the phases of atomic nuclei wave functions in numerical simulations.

In this work, we employ the transfer matrix (TM) method [40-43], an accurate numerical technique developed previously for calculating the probability of electrons tunneling through energy barriers [40], to study the NQEs of H atoms when they diffuse across the potential barriers/wells of Pt(111) surface. By means of the first-principles calculations, we determine the hydrogen diffusion paths on the Pt(111) surface and the according potential barriers. With moderate computational efforts, we apply the TM methods to describe the quantum effects of H diffusion on Pt(111), by taking the massive substrate atoms as classical particles. Resonant tunneling could occur when one H atom diffuses in the surface potential field where energy barriers and wells present alternately. It is shown that at room temperature and below, the quantum tunneling of hydrogen has significant effects on its surface diffusion. Further analysis on the temperature dependence of barrier-crossing reveals the existence of a nontrivial lower bound for the rate constant of surface diffusion and reactions even when temperature approaches the absolute zero.

The contents of this article are organized as follows: After Introduction, Section **2** introduces the TM method employed in this study and the technical details of the first-principles calculations. Section **3** presents the results of NQEs on the surface dynamics of H atoms on Pt(111), and analyzes in general the possible physical consequences due to the quantum tunneling of H. The main conclusion is summarized in Section **4**.

## 2. Formalism and Computational Methods

### 2.1 The Transfer Matrix Method

The transfer matrix (TM) method can be used to calculate the probability of a quantum particle passing through any types of potential field in one-dimensional situation. For the simplest case, a rectangular single potential barrier (well), only the boundary conditions which the particle experiences at the initial and final states need to be considered, and the corresponding Schrödinger equation is established. Each boundary gives a coefficient matrix which describes the amplitude of transmission and reflection of the wave function upon the transition. A transfer matrix which accounts for the transition between the initial and final states can be obtained by multiplying the coefficient matrices in order (see Appendix A). For a potential filed of arbitrary shape, as illustrated in Fig. 1, the basic idea is to divide the potential profile into a chain of slices, each of which can be regarded as a rectangular barrier (well).

As shown in Fig. 1, a particle propagates in the form of plane waves from the left, with the incident amplitude $A_R$, the reflection amplitude $B_L$, and the transmission amplitude $A_R$. Here $K_L$ is the incident wave vector and $K_R$ is the transmitted wave vector. The wave functions of the incident particle experience the reaction coordinates $x_0, x_1, \ldots, x_j, \ldots, x_N$, with the magnitude of potential at the corresponding points being $V_0, V_1, \ldots, V_j, \ldots, V_N$, respectively. The slices are of equal width $a$, i.e., $|x_{j+1} - x_j| = a$, for $1 \leq j \leq N$.

A coefficient matrix can be generated at each position that the particle passes. For instance, the coefficient matrix generated at the coordinate $x_j$ comes as follows:

$$M_j = \frac{1}{2k_j}\begin{pmatrix}(k_j + k_{j-1})e^{-(k_j - k_{j-1})x_j} & (k_j - k_{j-1})e^{-(k_j + k_{j-1})x_j} \\ (k_j - k_{j-1})e^{(k_j + k_{j-1})x_j} & (k_j + k_{j-1})e^{(k_j - k_{j-1})x_j}\end{pmatrix}, \quad (1)$$

where $k_{j-1} = \sqrt{2m(V_{j-1} - E)/\hbar^2}$, $k_j = \sqrt{2m(V_j - E)/\hbar^2}$, $x_j = (j-1)a$. The quantity $E$ is the particle incident energy, $m$ is the particle mass, and $\hbar$ is the reduced Planck's constant. In particular, for the incident potential point $V_0 = 0$ and $k_0 = \sqrt{2m(-E)/\hbar^2} = i\sqrt{2mE/\hbar^2}$, with $i$ being the imaginary unit. Then the chain product of $M_j$ gives the transfer matrix $M$ of the whole process as (Appendix A):

$$M = M_N M_{N-1} \ldots M_j \ldots M_2 M_1 \equiv \begin{pmatrix} m_{11} & m_{12} \\ m_{21} & m_{22} \end{pmatrix}. \quad (2)$$

The final transmission probability, i.e., the transmission coefficient may be calculated as follows (Appendix A):

$$T_r(E) = \left|\frac{A_R}{A_L}\right|^2 \times \frac{K_R}{K_L} = \frac{|M|^2}{|m_{22}|^2} \times \frac{K_R}{K_L}, \quad (3)$$

where $|M|$ is the determinant of transfer matrix $M$. For a system where the time-reversal symmetry presents, one has $K_R = K_L$, $|M| = 1$, and the transmission coefficient simplifies to (Appendix A) $T_r(E) = \frac{1}{|m_{22}|^2}$.

In practice, the matrix multiplication can be realized by numerical programming. The larger *N* is, the more accurate the resulting tunneling probability is. This is similar to the numerical evaluation of an integral: large values of *N* can in principle lead to numerical results which approach the exact value with arbitrary accuracies. Our calculations for barriers with a width of 1 Å show that, compared with $N = 100$ (i.e., sampling interval $\Delta x = 0.01$ Å), the case $N = 20$ ($\Delta x = 0.05$ Å) gives an averaged relative error of ~ 3%, and a correlation coefficient of ~ 0.9997 for the two $T_r(E)$ curves obtained, an indication of good convergence. In the case when $(V_j - E) < 0$ (incident energy is greater than the barrier height, or the potential well case), the indices $k_{j-1}$ and $k_j$ in the matrix $M_j$ of Eq. (1) become complex numbers, and the above expression still holds. Therefore, a unified treatment for the transmission of a quantum particle moving across a given potential filed is obtained using the TM method.

### 2.2 Details of First-principles Calculations

The first-principles calculations are carried out by the the Vienna *ab-initio* simulation package (VASP) [44, 45], which is based on density functional theory (DFT), to optimize the structure and calculate the ground state energies. The exchange correlation term is described by the generalized gradient approximation (GGA) with the PBE type functional [46], combined with the projector augmented wave (PAW) potentials [47, 48] to describe the electron−ion interactions. The energy cutoff of the plane wave basis set is 600 eV. The Pt(111) surface is modeled by a six-layer slab, with a *p*(3×3) surface unit cell which repeats periodically along the *xy* plane, and a vacuum layer of about 15 Å in the *z* direction. The bottom three layers of atomic coordinates are fixed to simulate the bulk phase, and the remained coordinates are released to simulate the surface phase. To eliminate the artificial dipole-dipole interactions caused by the upper and lower asymmetric slab surfaces, dipole corrections to the total energy are employed. A 4 ×4 ×1 Monkhorst-Pack k-mesh [49] is generated for sampling the Brillouin zone (BZ) with regard to the structural relaxation and total energy calculations. The transition states from one site to another site and the minimum energy paths (MEP) are obtained by using the nudged elastic band (NEB) method [50, 51]. The adsorption energy is: $E_{ads} = E[Pt(111)] + E[H] - E[H/Pt(111)] + \Delta E_{ZPV}$, where the terms $E[H/Pt(111)]$, $E[Pt(111)]$, $E[H]$ are the total energies of the system, the Pt(111) substrate and an isolated hydrogen atom, respectively. The last term $\Delta E_{ZPV} = \frac{1}{2}(\sum_{i,isolated} \hbar\omega_i - \sum_{j,ads} \hbar\omega_j)$, is the energy correction due to the change of the zero-point vibration energy of H atom from the isolated state to the surface adsorption state. In this work, $\Delta E_{ZPV} = -\frac{1}{2}\sum_{j,ads} \hbar\omega_j$, since $\omega_i \approx 0$ for a hydrogen atom in isolated state. The DFPT method is employed to calculate the vibrational frequencies [52].

To make a comparison with the results obtained by TM method, a semi-classical method, the Wentzel–Kramers–Brillouin (WKB) approximation [53] is also employed to study the quantum tunneling of H. For a given energy barrier $V(x)$, the WKB method computes the transmission coefficient as follows:

$$T_r(E) = \exp\left(-\frac{2}{\hbar}\int_a^b \sqrt{2m(V(x) - E)}\, dx\right), \qquad (4)$$

for $V(x) > E$ within the interval $a \leq x \leq b$, with $m$ being the particle mass; and $T_r(E) = 1$ when $V(x) \leq E$. In this work, $V(x)$ is determined using the NEB method.

### 3. Results and Discussions

*3.1 Adsorption and Diffusion of H on Pt(111).* We begin with investigating the adsorption and diffusion of individual H atoms on the Pt(111) surface. Then we applied the above method to study the quantum motions of H adatoms on Pt(111), which play an significant role in the anode reactions of fuel cells [54, 55]. Although the dynamics of H on Pt(111) have been extensively investigated [56-62], the understanding on quantum motions of H is still incomplete. For instance, depending on the method employed for experimental measurements, the reported diffusion coefficients of H on Pt(111) can differ by several orders of magnitude [57, 60-62]. Our first-principles calculations show that the total energy of molecular adsorption state of a single $H_2$ molecule is about 0.9 eV higher than that of the dissociated adsorption state on Pt(111). This implies that in the situation of low coverages, hydrogen exists in the form of atomic adsorption on the Pt(111), which is the system we are concerned with. First, we investigated three typical adsorption sites (top, fcc, hcp) of H on Pt(111). The relevant adsorption energies and geometric parameters are listed in Table 1. The magnitude of the adsorption energies indicates that H atoms are chemically adsorbed on Pt(111). For all the three configurations, the adsorption energies ($E_{ads}$) are ~ 2.6 eV, with a difference of less than 3%. On the other hand, adsorption on the top sites can be distinguished from the fcc and hcp sites by both zero-point energies (ZPE) and the H-Pt bond lengths, which are in good agreement with experimental data [56]. The calculated values of ZPE also compare well with previous work [59]. Before the correction of ZPE, the order of $E_{ads}$ is fcc > top > hcp, while it changes to fcc > hcp > top after ZPE correction. Such a change may have some minor modifications on the energy pathway of diffusion as discussed below.

Basically, one of the key factors governing the diffusion of surface adatoms is the minimum energy pathway (MEP) which joins the saddle points on the potential energy surface (PES). Here we used the NEB method to determine the MEP, i.e., the optimal paths for the diffusion of H atoms between typical adsorption sites as mentioned above. Figure 2 shows the MEP experienced by H atoms upon diffusion between adjacent isotype surface sites of Pt(111), i.e., fcc site to fcc site (labeled as fcc-fcc), hcp site to hcp site (labeled as hcp-hcp), and top site to top site (labeled as top-top). On the left panel of Fig. 2 are the potential energy curves, and on the right are the corresponding atomic configurations along the diffusion paths with the travelling distances of about 3 Å. Among them, the top-top barrier (~ 0.10 eV) is the highest and the hcp-hcp barrier (~ 0.006 eV) is the lowest. On the other hand, it can be seen that the energy profiles for the diffusion along paths hcp-hcp and top-top are not completely symmetric. This is due to the fact that diffusion paths are not perfectly symmetric in real space. For instance, near the midpoint of the path joining the two top sites as shown in Fig. 2, the left side is the hcp site while right side is the fcc site. The slight asymmetry is reflected in the potential energy surfaces of surface diffusion.

Figure 3 shows the calculated MEP for H diffusion between different types of adjacent surface sites (fcc-hcp, top-fcc, top-hcp) on Pt(111) surface. Shown on the left panels are the potential energy curves, and on the right are the atomic configurations along the optimal path of diffusion ($a \rightarrow b \rightarrow c \rightarrow d$). The length of each path is about 1.8 Å. Among them, the top-hcp barrier (~ 0.13 eV) is the highest and the fcc-hcp barrier is the lowest (~ 0.06 eV). The magnitudes of barriers height compare well with the results reported previously [58]. The diffusion pathways shown in Fig. 2 and Fig. 3 represent the elementary atomic processes of H diffusion on Pt(111), which can be regarded as the bases paths. Indeed, the diffusion between any two typical surface sites (fcc, top, hcp) at arbitrary separations is simply a linear combination of the diffusion pathways studied here. In addition, the low energy barrier of diffusion implies that low-coverage H atoms would be highly mobile on Pt(111) at room temperature and even below.

*3.2 The role of NQEs on H diffusion — Atomic Resonant Tunneling.* Then we turn to the studies of NQEs on the diffusion of H, with special attention paid to double barriers, which are the prototype of some important semiconducting heterostructures. Since the pioneering theoretical work by Tsu and Esaki [63] and their subsequent experimental verification in the GaAlAs-GaAs-GaAlAs system [64], the resonance transmission phenomenon of electrons in superlattice materials has also been observed in experiments [65, 66]. The physical conditions for resonance transmission in rectangular double barriers have been theoretically expounded [67]. The basic formula for the calculation of rectangular double barrier tunneling is [42]: $T_{double} = \frac{T^2}{|1+R\,e^{-i2(qb+qw+\phi_t)}|^2}$, where $T$ is the transmission coefficient and $R$ is the reflection coefficient for a single rectangular barrier, with $T + R = 1$; $q$ is the wave number of the incident plane wave when the barrier height is zero; finally, $\phi_t$ is the corresponding phase of the transmitted amplitude, and the geometric parameter $b$ is the barrier width and $w$ is the barrier spacing.

What about the situation when the much heavier H atoms passing through realistic single and double barriers which are usually irregular-shaped? It is known that [42] the transmission behavior of electrons is much different when transmitting across single and double barriers. We go further to demonstrate this point by investigating the diffusion of H atoms across realistic single and double barriers on Pt(111). Left panels of Fig. 4 show the potential energy curve, the diagram of transmission coefficient for the path top-hcp. It can be seen that for such a single barrier, the tunneling probability calculated by WKB is comparable with that given by the TM method for $E < 0.12$ eV. By contrast, the transmission probabilities differ significantly along the double barrier path (right panels of Fig. 4, top-hcp-top) when the incident energy is close to 0.12 eV, at which the resonant tunneling occurs.

The calculations above have assumed a plane wave nature of the incident particle. For the more general situation, the nuclear wave function of the incident particle may be expanded using the Fourier series: $\psi_N(x) = \sum_k a_k e^{ikx}$, with the coefficients $a_k$ subjected to the constraint $\sum_k |a_k|^2 = 1$. The expected value of kinetic energy is therefore $\langle E_k \rangle = \sum_k |a_k|^2 \left(\frac{\hbar^2 k^2}{2m}\right)$. The total probability of transmission is given by $P_{tot}(T) = \sum_k |a_k|^2 T_r(E_k) \sim \int_0^\infty \Gamma(E_k, T) T_r(E_k) dE_k$. The kinetic distribution function $\Gamma(E_k, T)$ is related to $a_k$ by $\sum_k |a_k|^2 = \int_0^\infty \Gamma(E_k, T) dE_k = 1$. Numerically, one has $\sum_k |a_k|^2 \approx \sum_k \Gamma(E_k, T) \Delta E_k$, and then $\Gamma(E_k, T) \approx \frac{|a_k|^2}{\Delta E_k}$. The coefficient $a_k$ is a function of temperature, since the true wave function which describes the quantum nature of the particle under consideration depends on the temperature of the system. The dependence of $a_k$ on temperature can be deduced within the framework of statistical mechanics. In a canonical ensemble, $|a_k|^2 = \sum_j \rho_{jk} = \frac{1}{Z} \sum_j e^{-\beta E_{jk}}$, where $\rho_{jk}$ is the probability of the $j$th quantum state (with the eigenenergy $E_{jk}$) having the kinetic energy of $E_k = \left(\frac{\hbar^2 k^2}{2m}\right)$, and $Z = \sum_s e^{-\beta E_s}$ is the partition function, with $E_s$ being the energy of the $s$th level; $\beta = \frac{1}{k_B T}$, $k_B$ is the Boltzmann constant and $T$ is the temperature. Physically, $\Gamma(E_k, T)$ satisfies the following condition: $\Gamma(0, T) = \Gamma(\infty, T) = 0$.

For a system where the kinetic energies (consequently the total energies) of single particles vary continuously, the total probability may be evaluated as

$$P_{tot}(T) = \int_0^\infty p(E,T) T_r(E) dE , \quad (5)$$

where the term $p(E,T) = 2\pi(\frac{1}{\pi k_B T})^{3/2} \sqrt{E} e^{-E/k_B T}$ is the kinetic energy distribution [68], which is suitable for the particles in thermal-equilibrium systems where the parabolic momentum-energy relation presents and scalar potentials dominate the interactions [Appendix B]. It can be shown that $p(E,T)$ satisfies that common condition for $\Gamma(E_k, T)$ as listed above. In particular, for a given $T$, the maximum of $p(E,T)$ sits at the energy point $E = \frac{1}{2} k_B T$.

In the case of H/Pt(111), the partition function describing the quantum motions of the adsorbed H is $Z_{tot} = Z_V Z_{trans}$, where $Z_V$ is the vibrational partition function, and $Z_{trans}$ is the translational partition function. Within the harmonic approximation, $Z_V = \frac{1}{1-e^{-\hbar\omega_V/k_B T}}$. The typical value of $\hbar\omega_V$ is ~ 0.3 eV (2 × ZPE, see Table 1), which means that $e^{-\hbar\omega_V/k_B T} \sim 0$ and $Z_V \sim 1$ for room temperature and below. Therefore, the translational partition function plays the major role, i.e., $Z_{tot} = Z_{trans}$. The translational motions are due to the coupling between H and the low-frequency phonons of the Pt substrate, for which the kinetic energies can be viewed as a continuous spectrum (Debye model). In this regard, the distribution of translational kinetic energies of the adsorbed H can be viewed as continuous, which can be approximately described by the function $p(E,T)$ as given above. Consequently, Eq. (5) is applied to calculate the total transmission probabilities of H across the potential fields experienced on Pt(111). The bottom panels of Fig. 4 show the total transmission probabilities at finite temperatures, which are evaluated using Eq. (5). For both single and double barriers, the total transmission probability increases monotonically with temperature. In the low temperature (low-$T$) region, there are only minor differences between the total transmission probabilities obtained by WKB and the TM method and the curves are almost identical for $T \leq 150$ K.

One of the key differences between the TM method and the WKB approximation is the phenomenon of resonant tunneling, which is absent in the latter. As shown in previous studies [67], resonant tunneling of electrons can be modulated by the phase factor describing the resonance, which is a function of barrier spacing. In fact, resonant tunneling can be regarded as the consequence of quantum interference between the incident and reflected particle wave functions in the region in-between the two barriers. To figure out the subtle changes in the total transmission probability due to resonant tunneling, we continue to study the transmission properties of H across the double barriers by varying spacing between the single barrier of top-hcp and its reverse process (hcp-top). Shown in Fig. 5(a), is an ideal case in which the distance between the two single barriers is extended by a platform with a width of 3 Å. Such a path is named as top-hcp-platform-hcp-top. A more realistic path (top-hcp-hcp-top) is depicted in Fig. 5(b), in which small potential fluctuations present along the intersection path hcp-hcp with a length of 3.2 Å. Figures 5(c) and 5(d) are respectively the diagrams showing the transmission probabilities along the path top-hcp-platform-hcp-top and top-hcp-hcp-top, and Figs. 5(e) and 5(f) give the corresponding total probabilities of transmission. It is worth noting that both top-hcp-platform-hcp-top and top-hcp-hcp-top have two obvious resonance peaks at energies well below the barrier height (~ 0.13 eV), which is different from the path top-hcp-top and a demonstration of the role of inter-barrier spacing. By comparing the data displayed in Figs. 5(e)-(f), one sees that it is difficult to distinguish the curves in the left panel from the right, which implies that the ideal path shown in Fig. 5(a) is a good approximation to the true diffusion path shown in Fig. 5(b). On the other hand, from both Fig. 5(e) and Fig. 5(f), one can see that despite the minor differences, the total transmission probabilities calculated by WKB capture main features of that by the TM method quite well for $T \leq 150$ K. Aside from the overall comparability in describing the physics of quantum tunneling, the coincidence of the TM and WKB calculations at the low-$T$ region also lies in the fact that the summation/integral over kinetic energy has smeared out the differences in the transmission coefficient ($T_r(E)$) at the few isolated energy points due to resonant tunneling.

On the other hand, the total probabilities given by WKB are larger than that of TM for $T \sim 150$ K and above (Figs. 5(e)-(f)), and the difference is enlarged with elevating temperatures. This is understandable by considering the following two facts: 1) The accurate transmission coefficient $T_r(E)$ given by the TM method oscillates around 1 while it is always 1 by WKB for $E \geq E_b$; 2) The H atoms will have higher probability of occupying the states with high kinetic energies which consequently lead to higher probability of full transmission.

The energy barriers adopted for studies above are obtained using DFT-NEB method, in which the effects of thermal excitations and entropy are neglected. Such an approximation is valid for low-$T$ conditions. For room temperature ($T \geq 300$ K) and even above, the thermal and entropy effects can be viewed as perturbations to the PES and the barrier heights. As shown in Appendix C, the perturbations of a sine-shaped potential field with a magnitude of ~ 0.04 eV will introduce slight modifications to the transmission coefficients of passing through the original barrier with a height of 0.2 eV. The minor variations will not change the main results presented here.

*3.3 The role of NQEs on barrier-crossing.* To highlight the quantum effects on transmission, we have further studied the behavior of barrier-crossing of a number of barriers at 300 K. For comparison, the total transmission probabilities are divided into two parts: the part contributed from tunneling in which the incident energy is lower than the barrier height ($E < E_b$) and the part from crossing ($E \geq E_b$), i.e., the full transmission region in semi-classical treatment such as WKB. Each part of the probability is defined as follows: Q($E < E_b$) = $\int_0^{E_b} P(E,T)T_r(E)dE$; Q($E \geq E_b$) = $\int_{E_b}^{\infty} P(E,T)T_r(E)dE$. The results are listed Table 2, for the TM and WKB method. One can see from Table 2 that for single barriers (rectangular, top-fcc, top-hcp), the numerically accurate transmission probabilities given by TM method are always smaller than the values calculated by WKB approximation. The situation changes in the case of double barriers (top-hcp-hcp-top). For the interval $E < E_b$, the transmission probability Q($E < E_b$) calculated by TM method is larger than that obtained by WKB, This is due to the resonant tunneling of H which present in the TM method while absent in the semi-classical WKB approximation (Fig. 5(d)). The underlying physics is that, the interference enhancement due to the particle wave functions is not considered in WKB. For $E \geq E_b$, classical particles pass completely ($T_r(E) = 1$) while quantum particles have some probability to be reflected backward. Therefore, the semi-classical method always predicts a higher probability than the TM method which is basically a full quantum description. In addition, both methods show that the transmission from incident energies which exceed the barrier height contributes more for the total transmission.

The phenomenon of resonant tunneling occurs not only when quantum particles travel across double barriers, but also across single potential wells. We go further to explore the NQEs on the diffusion dynamics of H when passing through surface potential wells. Shown in Fig. 6, is the MEP for the diffusion of H along the path hcp-fcc-hcp on Pt(111), a realistic potential field with a potential well with a depth of ~ 0.06 eV, sandwiched by two small identical barriers with a height of ~ 0.01 eV. The transmission coefficient $T_r(E)$ given by WKB increases exponentially with the incident energy $E$ and equals 1 when $E > E_b$ (Fig. 6(b)). In contrast, significant quantum oscillations are found in the $T_r(E)$ calculated by TM method. The first resonant tunneling peak appears at $E \sim 0.5$ meV (Fig. 6(b), inset), which corresponds to a temperature $T \sim 5$ K. The existence of resonance tunneling at such low particle energy leads to the emergence of a local peak in total transmission probability ($P_{tot}$). As a result, within the low-$T$ region of ~ 5 K to 30 K (Fig. 6(c)) $P_{tot}$ does not drop monotonically with decreasing temperature as predicted by WKB and that in double barriers, but jumps up instead and arrives at the peak position at ~ 5 K.

The anomalous temperature-dependence of $P_{tot}$ in the low-$T$ region along the diffusion path of hcp-fcc-hcp (Fig. 6(c)) points to its physical consequence on the low-$T$ diffusion of H atoms. Generally, the rate constant for diffusion follows the Van't Hoff-Arrhenius relation [69, 70]: $k = \nu e^{-E_b/(k_B T)}$, where $E_b$ is the activation energy, and $k_B$ and $T$ have the usual meanings; $\nu$ is the prefactor with the order of magnitude of typical vibrational frequencies of the adatom. The term $e^{-E_b/(k_B T)}$ is actually the total probability of barrier-crossing at thermal-equilibrium condition, i.e., the quantity $P_{tot}$ evaluated in this work. Therefore, anomalous behavior of the rate of diffusion in the low-$T$ region can be expected for H atoms along the hcp-fcc-hcp path on Pt(111). Figure 6(d) shows the rate constant of diffusion $k$ as a function of temperature, evaluated using a prefactor of $\nu = 1\times10^{12}$/s. It is clearly demonstrated that the rate constant increases with decreasing temperature for the interval of 5 K $\leq T \leq$ 30 K.

Using the total transmission probability $P_{tot}(T)$ calculated by TM method, we are able to deduce the site-to-site diffusion (hopping) rate constant as $k = \nu P_{tot}(T)$. The mean square displacement, $\langle d^2(t) \rangle$, can be related to the diffusion (hopping) rate constant $k$ [71], and the diffusion coefficient $D$ as $\langle d^2(t) \rangle = l^2 kt = 2Dt$, with $l$ being the site-to-site distance under consideration. It follows that the diffusion coefficient is given by $D(T) = \frac{l^2 k}{2} = l^2 \nu P_{tot}(T)/2$. With a given prefactor $\nu = 10^{12} s^{-1}$, the diffusion coefficient $D(T)$ for the path top-hcp is 7.05×10$^{-10}$ cm$^2$s$^{-1}$, 1.05×10$^{-8}$ cm$^2$s$^{-1}$, and 6.45×10$^{-8}$ cm$^2$s$^{-1}$ at $T$ = 90 K, 120 K and 150 K, respectively. For diffusion along the path hcp-fcc-hcp, the value of $D(T)$ is respectively 2.63×10$^{-4}$ cm$^2$s$^{-1}$, 3.12×10$^{-4}$ cm$^2$s$^{-1}$, 3.47×10$^{-4}$ cm$^2$s$^{-1}$ at $T$ = 90 K, 120 K and 150 K. On the experimental side, available data on the diffusion coefficients of H on Pt(111) can be grouped into two categories based on the method employed: The data measured using the linear optical diffraction (LOD) technique [60, 61] and the data measured based on the technique of helium atom scattering (HAS) [57, 62]. For $T$ = 90 ~ 150 K, the calculated diffusion coefficients along the *path top-hcp* are slightly larger than that obtained by LOD (~ 10$^{-11}$ (90 K) to ~ 10$^{-8}$ cm$^2$s$^{-1}$ (150 K)) [60, 61], and are about two orders of magnitude smaller than the data given by HAS (~ 10$^{-6}$ (~ 90 K) to 10$^{-5}$ cm$^2$s$^{-1}$ (~ 150 K)) [57, 62]. In contrast, the diffusion coefficients along the *path hcp-fcc-hcp* are much higher than the values reported by both LOD and HAS measurements. In experimental measurements, the obtained diffusion coefficient $D(T)_{expt}$ is actually a statistical average of all possible diffusion paths, which therefore sets the interval of $D(T)_{top-hcp} < D(T)_{expt} < \frac{l^2}{2} \times 10^{-4}$ cm$^2$s$^{-1}$, with $P_{tot}(T) = 1$ for the upper bound, and the distance $l$ in units of Å. The difference in the diffusion coefficients given by LOD and HAS may arise from the spatial scale of Pt(111) involved in the measurements: The former spans multiple atomic terraces by crossing a number of atomic steps which are expected to have nontrivial effects on H diffusion, while the latter correspond to single hops on well-defined atomic terraces. In this regard, the model employed in our work is more likely to simulate the situation of HAS experiments [57, 62]. Indeed, it is clear that the data (~ $10^{-7}$ to $10^{-5}$ cm$^2$s$^{-1}$) obtained by HAS [57, 62] fall into the interval given above.

*3.4 The role of NQEs on low-temperature reaction rate.* Another notable physical consequence due to NQEs is the decrease of energy barriers that H atoms actually experienced within the framework of classical theory of reaction rate. The Van't Hoff-Arrhenius type relation for rate constant not only applies to the process of atomic surface diffusion, but also to a wide variety of chemical and physical processes [69]. As mentioned above, the exponential term $e^{-E_b/(k_B T)}$ is physically the total probability of passing through a given barrier. The effective barrier that H atoms actually experienced, $E_b^*$, can therefore be numerically deduced using the equality $P_{tot}(T) = e^{-E_b^*/(k_B T)}$, and one has $E_b^* = -(k_B T)\ln[P_{tot}(T)]$. The diffusion of H from top site to the fcc site of Pt(111) is taken as an example, to investigate the role of NQEs on barrier-lowering. The results are shown in Fig. 7, calculated for temperatures up to 500 K. The original data obtained using NEB method is a classical barrier in which the H atoms are treated classical particles within the adiabatic approximation, and the barrier height is reduced by ~ 0.04 eV after zero-point energy corrections. Using the TM method, total transmission probabilities are calculated for both barriers and the corresponding values of effective barriers are plotted in Fig. 7. It is obvious that the effective barrier is significantly lowered with comparison to the classical one. The effect is even more pronounced in the low-*T* region. For instance, at *T* ~ 10 K the effective barrier $E_b^*$ is nearly one order of magnitude smaller than the classical one. Another interesting feature as seen from Fig. 7 is that, $E_b^*$ increases quickly with temperature and arrives at its maximum at *T* ~ 100 K, and then deceases monotonically but smoothly with increasing $T$.

How to understand the appearance of the maximum of $E_b^*$? The TM method is numerically accurate; however, the variation trend with temperature is not easily explained from the data due to the absence of a simple analytic description. As demonstrated by the studies above, in the case of single barrier, WKB captures the main features of TM excellently for $E < E_b$. The advantage of WKB is that the transmission coefficient $T_r(E)$ has an analytic form. In the following, we attempt to explain the temperature dependence of $E_b^*$ within the WKB approximation. Mathematically, the expression of Eq. (5) may be rewritten as

$$P_{tot}(T) = \int_0^\infty P(E,T)T_r(E)dE = \lim_{E_m \to \infty} \int_0^{E_m} P(E,T)T_r(E)dE. \quad (6)$$

In practice, the numerical results of $P_{tot}(T)$ is well converged when the upper bound $E_m$ is two times larger than that of the barrier height, i.e., $E_m \geq 2E_b$. In particular, $E_m \sim E_b$ is sufficient to describe the low temperature transmission across single barriers. That is, the total probability $P_{tot}(T) \approx \int_0^{E_b} P(E,T)T_r(E)dE$. Using the mean value theorem for integrals, there exits one energy point $\xi$ ($0 < \xi = \lambda k_B T < E_b$) such that $P_{tot}(T) = T_r(\xi)\int_0^{E_b} P(E,T)dE$, where $\lambda$ is dimensionless. Recalling the constraint that the integral $\int_0^{E_b} P(E,T)dE \sim 1$, one has $P_{tot}(T) \approx T_r(\lambda k_B T)$, and consequently the effective barrier $E_b^* = -(k_B T)\ln[P_{tot}(T)] = -(k_B T)\ln[T_r(\lambda k_B T)]$. Within the WKB approximation which is good for $T_r(E)$ at low temperature, the effective barrier is calculated as follows:

$$E_b^* = \frac{2k_B T}{\hbar}\int_a^b \sqrt{2m(V(x) - \lambda k_B T)}dx. \quad (7)$$

Again, using the mean value theorem for integrals, one arrives at the following expression:

$$E_b^* = \frac{2k_B T(b-a)}{\hbar} \times \sqrt{2m(V(\eta) - \lambda k_B T)}, \quad (8)$$

where $a < \eta < b$, and $V(\eta) - \lambda k_B T \geq 0$. The first term $\frac{2k_B T(b-a)}{\hbar}$ increases with temperature $T$ while the second term $\sqrt{2m(V(\eta) - \lambda k_B T)}$ decreases with $T$. Compromise of these two competing terms results in the appearance of a maximum, which locates at $T_m = \frac{2V(\eta)}{3\lambda k_B}$. The maximum is therefore determined as $E_{bm}^* = \frac{2k_B T_m (b-a)}{\hbar} \times \sqrt{\frac{2}{3} mV(\eta)}$. Using the data presented in Fig. 7, the two parameters $V(\eta)$ and $\lambda$ in Eq. (8) can be fixed. The results are summarized in Table 3, for the maxima of the two $E_b^*$ curves shown in Fig. 7.

We note here that the classical barrier and its ZPE-corrected counterpart are taken as constant since the temperature effects on the potential energy surface and consequently the MEP are usually negligible for systems in which the thermal expansion is small (for Pt, $\alpha \sim 1\times 10^{-5}$/K [72]) within the temperature range considered, and the errors introduced by temperature would largely cancel for the different surface sites.

Another interesting property is the low-temperature behavior of the effective barrier. From Eq. (8) one sees the effective barrier $E_b^*$ would approaches 0 when the temperature approaches the absolute zero ($T \to 0$). Moreover, from the relation $E_b^* = \frac{2k_B T(b-a)}{\hbar} \times \sqrt{2m(V(\eta) - \lambda k_B T)} < \frac{2k_B T(b-a)}{\hbar} \times \sqrt{2mV(\eta)} = E_{bup}^*$, which is the upper bound of $E_b^*$, one can understand the nearly linear decay of $E_b^*$ in the low-$T$ region (Fig. 7). In the situation when the Van't Hoff-Arrhenius relation applies, for instance, a surface-based diffusion or reaction process, the lower bound (low-$T$ limit) of rate constant (when $T \to 0$) can therefore be established:

$$k(T \to 0) = \nu e^{-E_b^*/(k_B T)} = \nu e^{-\frac{2(b-a)}{\hbar} \times \sqrt{2mV(\eta)}} \geq \nu e^{-\frac{2(b-a)}{\hbar} \times \sqrt{2mE_b}}. \qquad (9)$$

More generally, the low-$T$ limit expressed in Eq. (9) can be rewritten as

$$k(T \to 0) \geq \nu e^{-\frac{2L}{\hbar} \times \sqrt{2mE_b}} \equiv k_{min}, \qquad (10)$$

where $L = (b - a)$ is the distance travelled by the particle in the one-dimensional or quasi-one-dimensional potential field, $E_b$ is the classical barrier height, and the other parameters have the usual meanings as above. It should be stressed here that, the existence of such a lower bound does not depend on the detailed formulation of the kinetic energy distribution function; the only precondition is the WKB approximation, which is shown to be valid for barrier-crossing process in low-$T$ region.

From Eq. (10) it is clear that $k_{min}$ decreases quickly with distance $L$, particle $m$

and barrier height $E_b$, and is independent of temperature. The existence of nonzero lower bound of rate constant naturally points to the possibility of active surface diffusion and chemical reactions involving H atoms even at ultra-low temperatures. The effect would be pronounced for the systems where small barriers dominate the key processes. This prediction is in contrast to the general empirical intuitions that diffusion and reactions of atoms on surfaces tend to cease when the temperature is close to the absolute zero. For instance, at $T = 1$ K, the effective barrier for diffusion along the path top-fcc is $E_b^* = 3.0299 \times 10^{-5}$ eV. The corresponding parameters are: $L \sim 1.84$ Å, $\lambda \sim 1247$, $V(\eta) \sim 0.1079$ eV. Given that the prefactor $\nu \sim 10^{12}$/s, one has $k(T \to 0) \sim 2.767$/s, and $k_{min} \sim 2.138$/s. Such a rate constant implies that the diffusion process can still be fairly active even at cryogenic conditions.

## 4. Conclusions

To summarize, the nuclear quantum effects (NQEs) on the diffusion dynamics of H have been studied using the technique of transfer matrix (TM), which is numerically accurate for describing the transport behavior of quantum particles in one-dimensional (or mathematically equivalent) potential fields. The atomic resonant tunneling (ART) of H atoms is demonstrated in realistic potentials which describe the diffusion of H on Pt(111). In addition, the significance of phase factor in resonant tunneling of double barriers is revealed by investigating the dependence of tunneling probability with the inter-barrier spacing. Along the diffusion paths which connect the typical surface sites of Pt(111), the energy barriers calculated by DFT within the adiabatic approximation turn out to be ~ 0.1 eV, which indicates the high mobility of H on Pt(111) surface. Owing to resonant tunneling, the transmission and the rate constant of diffusion along the path hcp-fcc-hcp is predicted to increase abnormally with decreasing temperature within the interval 5 K $\leq T \leq$ 30 K. Based on the probabilities of barrier-crossing calculated at finite temperature, the concept of effective barrier ($E_b^*$) is introduced for Van't Hoff-Arrhenius type processes and analyzed for some typical diffusion path. The temperature-dependent curves of effective barrier show a maximum at certain temperature $T_m$, below or above which $E_b^*$ drops monotonically with varying temperature. The nearly linear decay of effective barrier with temperature $T$ results in nontrivial rate constant even when $T \rightarrow 0$. The vanishing effective barrier consequently makes active surface diffusion and reactions involving H possible at ultra-low temperatures. Analysis based on WKB approximation yield a nonzero low-$T$ limit for rate constant. The results presented here are expected to be tested by low-$T$ reactions involving H and its isotope D, for which the reactions rates are expected to be distinguished experimentally.

**Acknowledgements**

This work is financially supported by the National Natural Science Foundation of China (No. 11474285). We would like to thank Professor Wang Enge and Professor Zhu Wenguang for their reading and helpful comments on the manuscript. We also thank Mr. Wang Jiawei for his helpful discussions on the general properties of transfer matrix. We gratefully acknowledge the high-performance supercomputing service from AM-HPC, and the staff of the Hefei Branch of Supercomputing Center of Chinese Academy of Sciences for their support of the supercomputing facilities.

**Appendix A**

In this appendix, we provide the mathematical details on the transfer matrix (TM) method, for the calculation of transmission probability (transmission coefficient) of quantum particles passing through a given potential field.

### A.1 Single rectangular barrier

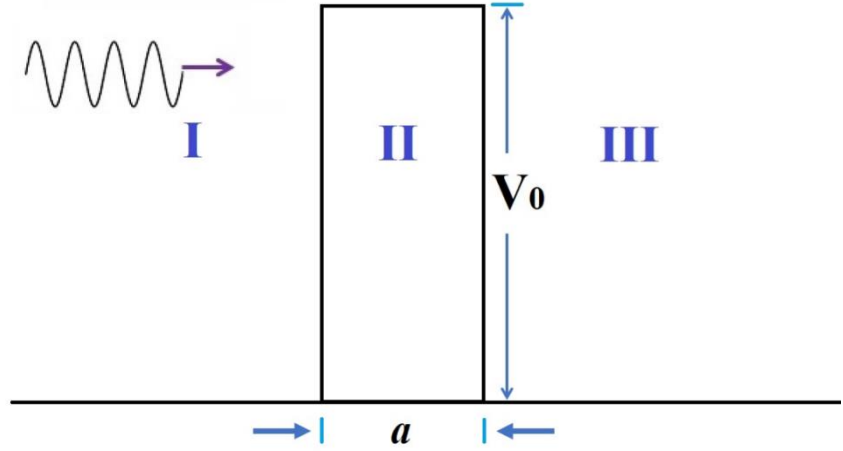

**Figure A1** Schematic diagram for a quantum particle passing through a rectangular barrier.

As schematically shown in Fig. A1, a quantum particle with incident energy $E$ passes through a rectangular barrier with the height of $V_0$ and width $a$. The Schrödinger equation for this process reads:

$$\hat{H}\psi = E\psi, \qquad (A1)$$

with $\hat{H} = \frac{-\hbar^2}{2m}\nabla^2 + V(x)$; $V(x) = V_0$ for $0 \leq x \leq a$, and $V(x) = 0$ otherwise; $m$ is the particle mass and $\hbar$ is the reduced Planck's constant.

The wave functions in regions I, II, III may be written as:

$$\begin{aligned}
\psi_I &= A_L e^{ikx} + B_L e^{-ikx} & (x<0), & \quad k = \sqrt{2mE/\hbar^2}. \\
\psi_{II} &= C e^{\beta x} + D e^{-\beta x} & (0 \leq x \leq a), & \quad \beta = \sqrt{2m(V_0 - E)/\hbar^2}. \quad (A2)\\
\psi_{III} &= A_R e^{ikx} + B_R e^{-ikx} & (x>a).
\end{aligned}$$

Making advantage of the continuity of wave function (wave function match) and its derivative at the boundaries, for $x = 0$ one has

$\psi_I(0) = \psi_{II}(0)$, $\frac{d\psi_I}{dx}\big|_{x=0} = \frac{d\psi_{II}}{dx}\big|_{x=0}$, and then arrives at

$$\begin{cases} A_L + B_L = C + D \\ ik(A_L - B_L) = \beta(C - D) \end{cases} \Rightarrow \begin{cases} C = \frac{1}{2}[\frac{\beta+ik}{\beta}A_L + \frac{\beta-ik}{\beta}B_L] \\ D = \frac{1}{2}[\frac{\beta-ik}{\beta}A_L + \frac{\beta+ik}{\beta}B_L] \end{cases} \quad (A3)$$

The results in Eq. (A3) can be expressed in the matrix form:

$$\begin{pmatrix} C \\ D \end{pmatrix} = \frac{1}{2\beta}\begin{pmatrix} \beta + ik & \beta - ik \\ \beta - ik & \beta + ik \end{pmatrix}\begin{pmatrix} A_L \\ B_L \end{pmatrix} \quad (A4)$$

Again, using the wave function match at $x = a$, $\psi_{II}(a) = \psi_{III}(a)$ $\frac{d\psi_{II}}{dx}|_{x=a} = \frac{d\psi_{III}}{dx}|_{x=a}$, one has

$$\begin{cases} Ce^{\beta a} + De^{-\beta a} = A_R e^{ika} + B_R e^{-ika} \\ \beta(Ce^{\beta a} - De^{-\beta a}) = ik(A_R e^{ika} - B_R e^{-ika}) \end{cases} \Rightarrow$$

$$\begin{pmatrix} A_R \\ B_R \end{pmatrix} = \frac{1}{2ik}\begin{pmatrix} (ik+\beta)e^{-(ik-\beta)a} & (ik-\beta)e^{-(ik+\beta)a} \\ (ik-\beta)e^{(ik+\beta)a} & (ik+\beta)e^{(ik-\beta)a} \end{pmatrix}\begin{pmatrix} C \\ D \end{pmatrix} \quad (A5)$$

Combining Eq. (A4) and Eq. (A5), one has

$$\begin{pmatrix} A_R \\ B_R \end{pmatrix} = \frac{1}{2ik}\begin{pmatrix} (ik+\beta)e^{-(ik-\beta)a} & (ik-\beta)e^{-(ik+\beta)a} \\ (ik-\beta)e^{(ik+\beta)a} & (ik+\beta)e^{(ik-\beta)a} \end{pmatrix}\frac{1}{2\beta}\begin{pmatrix} \beta + ik & \beta - ik \\ \beta - ik & \beta + ik \end{pmatrix}\begin{pmatrix} A_L \\ B_L \end{pmatrix}$$

(A6)

Let $M = \frac{1}{2ik}\begin{pmatrix} (ik+\beta)e^{-(ik-\beta)a} & (ik-\beta)e^{-(ik+\beta)a} \\ (ik-\beta)e^{(ik+\beta)a} & (ik+\beta)e^{(ik-\beta)a} \end{pmatrix}\frac{1}{2\beta}\begin{pmatrix} \beta + ik & \beta - ik \\ \beta - ik & \beta + ik \end{pmatrix}$, then

The incident and outgoing amplitudes of wave functions are expressed as follows:

$$\begin{pmatrix} A_R \\ B_R \end{pmatrix} = M\begin{pmatrix} A_L \\ B_L \end{pmatrix} \equiv \begin{pmatrix} m_{11} & m_{12} \\ m_{21} & m_{22} \end{pmatrix}\begin{pmatrix} A_L \\ B_L \end{pmatrix} \quad (A7)$$

with $M = \begin{pmatrix} m_{11} & m_{12} \\ m_{21} & m_{22} \end{pmatrix}$, which is commonly referred to as transfer matrix.

Once the transfer matrix is known, the transmission properties would be fully understood.

## A.2 Generalization to a potential field of arbitrary shape

In this subsection, we extend the above deduction of transfer matrix to the more generalized situation, a given potential field $V(x)$ where potential barriers and/or wells present, by using the condition of wave function match. For a quantum particle in a given potential field $V(x)$, the Schrödinger Equation:

$$-\frac{\hbar^2}{2m}\frac{d^2}{dx^2}\psi + V(x)\psi = E\psi \qquad (A8)$$

That is,

$$\frac{d^2}{dx^2}\psi = [2m(V(x) - E)/\hbar^2]\psi \qquad (A9)$$

Let $k^2 = 2m(V(x) - E)/\hbar^2$, $k = \sqrt{2m(V(x) - E)/\hbar^2}$, $k$: complex number.

The solution of Eq. (A9) can be generally expressed as follows:

$$\psi(x) \sim Ae^{kx} + Be^{-kx} \qquad (A10)$$

Numerically, the diffusion path can be divided into ($N$+1) parts/regions with equal width and magnitude $V_n$, and $N$ boundary lines, then for the $n$th region:

$$\psi_n(x) \sim A_n e^{k_n x} + B_n e^{-k_n x}, \ k_n = \sqrt{2m(V_n - E)/\hbar^2}, n = 0, 1, 2, \ldots N \qquad (A11)$$

Using the condition of wave function match ($\psi_n = \psi_{n+1}$; $\psi'_n = \psi'_{n+1}$) at $x_{n+1}$, one has

$$\begin{cases} A_n e^{k_n x_{n+1}} + B_n e^{-k_n x_{n+1}} = A_{n+1} e^{k_{n+1} x_{n+1}} + B_{n+1} e^{-k_{n+1} x_{n+1}} \\ k_n A_n e^{k_n x_{n+1}} - k_n B_n e^{-k_n x_{n+1}} = k_{n+1} A_{n+1} e^{k_{n+1} x_{n+1}} - k_{n+1} B_{n+1} e^{-k_{n+1} x_{n+1}} \end{cases}$$

$$(A12)$$

The coefficients $A_{n+1}$, $B_{n+1}$ are

$$\begin{cases} A_{n+1} = \frac{1}{2k_{n+1}}[(k_{n+1} + k_n)A_n e^{-(k_{n+1} - k_n)x_{n+1}} + (k_{n+1} - k_n)B_n e^{-(k_{n+1} + k_n)x_{n+1}}] \\ B_{n+1} = \frac{1}{2k_{n+1}}[(k_{n+1} - k_n)A_n e^{(k_{n+1} + k_n)x_{n+1}} + (k_{n+1} + k_n)B_n e^{(k_{n+1} - k_n)x_{n+1}}] \end{cases}$$

(A13)

Eq. (A13) can be expressed in the matrix form:

$$\begin{pmatrix} A_{n+1} \\ B_{n+1} \end{pmatrix} = \frac{1}{2k_{n+1}} \begin{pmatrix} C_{n+1}e^{-D_{n+1}x_{n+1}} & D_{n+1}e^{-C_{n+1}x_{n+1}} \\ D_{n+1}e^{C_{n+1}x_{n+1}} & C_{n+1}e^{D_{n+1}x_{n+1}} \end{pmatrix} \begin{pmatrix} A_n \\ B_n \end{pmatrix} \qquad (A14)$$

where $C_{n+1} = k_{n+1} + k_n$, $D_{n+1} = k_{n+1} - k_n$

The $j$th transition matrix is therefore,

$$M_j = \frac{1}{2k_j}\begin{pmatrix} (k_j + k_{j-1})e^{-(k_j - k_{j-1})x_j} & (k_j - k_{j-1})e^{-(k_j + k_{j-1})x_j} \\ (k_j - k_{j-1})e^{(k_j + k_{j-1})x_j} & (k_j + k_{j-1})e^{(k_j - k_{j-1})x_j} \end{pmatrix}. \qquad (A15)$$

The chain product of $M_j$ gives the total transfer matrix $M$.

**A.3 General properties of transfer matrix**

In this subsection, we study the general properties of transfer matrix, which relates the incoming and outgoing amplitudes of wave functions.

For a quantum particle propagates from left to right (Fig. 1), the wave function

$$\begin{cases} \psi_L = A_L e^{ik_L x} + B_L e^{-ik_L x} \\ \psi_R = A_R e^{ik_R x} + B_R e^{-ik_R x} \end{cases} \tag{A16}$$

The amplitudes of wave functions are related by the transfer matrix as follows:

$$\begin{pmatrix} A_R \\ B_R \end{pmatrix} = M \begin{pmatrix} A_L \\ B_L \end{pmatrix} = \begin{pmatrix} m_{11} & m_{12} \\ m_{21} & m_{22} \end{pmatrix} \begin{pmatrix} A_L \\ B_L \end{pmatrix} \tag{A17}$$

The probability current reads

$$J(x) = -\frac{i\hbar}{2m}\left[\psi^* \frac{d}{dx}\psi - \psi \frac{d}{dx}\psi^*\right], \quad \text{and} \quad \frac{d}{dx}J(x) = -\frac{i\hbar}{2m}\left[\psi^* \frac{d^2}{dx^2}\psi - \psi \frac{d^2}{dx^2}\psi^*\right] = 0$$

(A18)

It follows that $J(x) =$ constant.

This is the conservation of probability current.

In particular,

$$J_L = \frac{\hbar K_L}{m}(|A_L|^2 - |B_L|^2) = J_R = \frac{\hbar K_R}{m}(|A_R|^2 - |B_R|^2). \tag{A19}$$

For the case shown in Fig. 1, $B_R = 0$. From Eq. (A17), one has

$$\begin{cases} A_R = A_L m_{11} + B_L m_{12} \\ 0 = A_L m_{21} + B_L m_{22} \end{cases} \Rightarrow A_R = \left(\frac{m_{11}m_{22} - m_{12}m_{21}}{m_{22}}\right) A_L \tag{A20}$$

Meanwhile, the transmission coefficient $T_r(E)$ is defined as

$$T_r(E) = \frac{J_{out}}{J_{in}} = \frac{\frac{\hbar K_R}{m}|A_R|^2}{\frac{\hbar K_L}{m}|A_L|^2} = \frac{|A_R|^2}{|A_L|^2} \times \frac{K_R}{K_L} \tag{A21}$$

In combination with Eq. (A20), one arrives at

$$T_r(E) = \frac{|A_R|^2}{|A_L|^2} \times \frac{K_R}{K_L} = \frac{|M|^2}{|m_{22}|^2} \times \frac{K_R}{K_L}, \tag{A22}$$

where $|M| = m_{11}m_{22} - m_{12}m_{21}$, is the determinant of $M$.

In the case when time-reversal symmetry presents, i.e., the effective Hamiltonian

$\widehat{H}$(left → right) = $\widehat{H}$(right → left) = $\widehat{H}$ in Fig. 1, $K_R = K_R = k$, with the same eigenvalue $E$. From the complex conjugate $\widehat{H}\psi^* = E\psi^*$, one has the following

$$\begin{cases} \psi_L^* = A_L^* e^{-ikx} + B_L^* e^{ikx} \\ \psi_R^* = A_R^* e^{-ikx} + B_R^* e^{ikx} \end{cases} \Rightarrow \begin{cases} \psi_L^* = B_L^* e^{ikx} + A_L^* e^{-ikx} \\ \psi_R^* = B_R^* e^{ikx} + A_R^* e^{-ikx} \end{cases} \quad (A23)$$

$$\begin{pmatrix} B_R^* \\ A_R^* \end{pmatrix} = \begin{pmatrix} m_{11} & m_{12} \\ m_{21} & m_{22} \end{pmatrix} \begin{pmatrix} B_L^* \\ A_L^* \end{pmatrix} \quad (A24)$$

$$\begin{pmatrix} A_R \\ B_R \end{pmatrix} = \begin{pmatrix} m_{22}^* & m_{21}^* \\ m_{12}^* & m_{11}^* \end{pmatrix} \begin{pmatrix} A_L \\ B_L \end{pmatrix} \quad (A25)$$

By comparison with $\begin{pmatrix} A_R \\ B_R \end{pmatrix} = \begin{pmatrix} m_{11} & m_{12} \\ m_{21} & m_{22} \end{pmatrix} \begin{pmatrix} A_L \\ B_L \end{pmatrix}$, it is easily found that $m_{11} = m_{22}^*$, $m_{12} = m_{21}^*$.

On the other hand, the equality in Eq. (A19) reduces to

$$J_L = \frac{\hbar k}{m}(|A_L|^2 - |B_L|^2) = J_R = \frac{\hbar k}{m}(|A_R|^2 - |B_R|^2) \quad (A26)$$

Then

$$|A_L|^2 - |B_L|^2 = |A_R|^2 - |B_R|^2 \quad (A27)$$

Substitution using the relations established previously: $\begin{cases} A_R = m_{11}A_L + m_{12}B_L \\ B_R = m_{21}A_L + m_{22}B_L \end{cases}$ & $\begin{cases} m_{11} = m_{22}^* \\ m_{12} = m_{21}^* \end{cases}$, one has the following:

$$|A_R|^2 - |B_R|^2 = |m_{11}A_L + m_{22}B_L|^2 - |m_{21}A_L + m_{22}B_L|^2 =$$
$$(|m_{11}|^2 - |m_{21}|^2)(|A_L|^2 - |B_L|^2) \quad (A28)$$

Consequently,

$$|M| = \det M = |m_{11}|^2 - |m_{21}|^2 = 1 \quad (A29)$$

The transmission coefficient simplifies to

$$T_r(E) = \frac{1}{|m_{22}|^2} = \frac{1}{|m_{11}|^2} \quad (A30)$$

## Appendix B

In this appendix, we study the distribution of kinetic energy of single particles in a scalar potential field. As shown by the previous work of one of the authors [68], provided that the velocity of a single particle in a thermal-equilibrium many-particle system obeys the Maxwell velocity distribution:

$$f(v) = 4\pi \left(\frac{m}{2\pi k_B T}\right)^{3/2} v^2 e^{-\frac{mv^2}{2k_B T}}, \qquad (B1)$$

then the distribution of kinetic energy of the single particle may be expressed as follows [68]:

$$p(E_k) = 2\pi \left(\frac{1}{\pi k_B T}\right)^{3/2} \sqrt{E_k} e^{-\frac{E_k}{k_B T}}. \qquad (B2)$$

Here, we show in general, the Maxwell velocity distribution applies to canonical ensembles (NVT) in which the interactions between the constituent particles are described by scalar potentials, i.e., the strength of interactions depends only on the particle-particle separations. With this precondition, for a canonical ensemble which consists of $N$ particles with a volume $V$ at temperature $T$, the total energy $E$ of the system may be written as:

$$E = \sum_{i=1}^{N} \frac{\vec{p}_i^2}{2m_i} + \sum_{i<j} V(q_{ij}), \qquad (B3)$$

where $\vec{p}_i$ is the momentum of the $i$th particle with mass $m_i$; $V(q_{ij})$ is the interaction potential between the $i$th and $j$th particle, with the $q_{ij} = |\vec{q}_i - \vec{q}_j|$, i.e., the separation between the two particles with coordinates $\vec{q}_i$ and $\vec{q}_j$. Given that the momenta of single particles vary continuously, then the partition function of the system reads:

$$Z = \frac{1}{\prod_j N_j! h^{3N}} \int \cdots \int e^{-\beta E} d\vec{q}_1 d\vec{q}_2 \cdots d\vec{q}_N d\vec{p}_1 d\vec{p}_2 \cdots d\vec{p}_N \equiv \frac{1}{\prod_j N_j! h^{3N}} \int e^{-\beta E} d\Omega, \quad (B4)$$

where the differential elements $d\vec{q}_j = dx_j dy_j dz_j$, $d\vec{p}_j = dp_{xj} dp_{yj} dp_{zj}$; $\sum_j N_j = N$, where $N_j$ is the number of $j$th type particle; and the quantity $d\Omega = d\vec{q}_1 d\vec{q}_2 \cdots d\vec{q}_N d\vec{p}_1 d\vec{p}_2 \cdots d\vec{p}_N$, is the differential volume element of the coordinates-momenta phase space; $\beta = \frac{1}{k_B T}$; $h$ is the Planck's constant. For simplicity of discussion, let $q = (\vec{q}_1, \vec{q}_2, \ldots, \vec{q}_N)$, $p = (\vec{p}_1, \vec{p}_2, \ldots, \vec{p}_N)$. The probability of finding the microscopic state $(q, p)$ in volume element $d\Omega$ is:

$$\rho(q,p) d\Omega = \frac{1}{\prod_j N_j! h^{3N}} \frac{e^{-\beta E(q,p)}}{Z} d\Omega. \qquad (B5)$$

For the $n$th particle with a mass of $m_n$, and with the coordinate $\vec{q}_n = (x_n, y_n, z_n)$, and momentum $\vec{p}_n = (p_{xn}, p_{yn}, p_{zn})$, the probability of finding this particle in the momentum range $\vec{p}_n \to \vec{p}_n + d\vec{p}_n$ is given by

$$\rho(\vec{p}_n)d\vec{p}_n = \frac{1}{\prod_j N_j! h^{3N}} \frac{e^{-\beta E_{kn}}}{Z_n} d\vec{p}_n \frac{\int e^{-\beta \tilde{E}} d\tilde{\Omega}}{\tilde{Z}}, \qquad (B6)$$

where $E_{kn} = \frac{\vec{p}_n^2}{2m_n}$, is the kinetic energy of the $n$th particle, $\tilde{E} = E - E_{kn}$, $d\tilde{\Omega} = \frac{d\Omega}{d\vec{p}_n}$, and $Z_n = \frac{1}{N_n! h^3} \int e^{-\beta E_{kn}} d\vec{p}_n$, $\tilde{Z} = \frac{N_n!}{\prod_j N_j h^{3(N-1)}} \int e^{-\beta \tilde{E}} d\tilde{\Omega} = \frac{Z}{Z_n}$.

It follows that,

$$\rho(\vec{p}_n)d\vec{p}_n = \frac{e^{-\beta E_{kn}}}{\int e^{-\beta E_{kn}} d\vec{p}_n} d\vec{p}_n. \qquad (B7)$$

Recalling that $E_{kn} = \frac{\vec{p}_n^2}{2m_n} = \frac{p_{nx}^2 + p_{ny}^2 + p_{nz}^2}{2m_n}$, $d\vec{p}_n = dp_{nx} dp_{ny} dp_{nz}$, the integral turns out to be $\int e^{-\beta E_{kn}} d\vec{p}_n = (2\pi m k_B T)^{3/2}$.

In the spherical coordination system, the term $d\vec{p}_n$ transforms to $d\vec{p}_n = 4\pi p_n^2 dp_n$, with $p_n = \sqrt{p_{nx}^2 + p_{ny}^2 + p_{nz}^2} = m_n v_n$, and consequently one has

$$d\vec{p}_n = 4\pi p_n^2 dp_n = 4\pi m_n^3 v_n^2 dv_n. \qquad (B8)$$

Then Eq. (B7) reduces to

$$\rho(\vec{p}_n)d\vec{p}_n = 4\pi \left(\frac{m}{2\pi k_B T}\right)^{3/2} v_n^2 e^{-\frac{m v_n^2}{2 k_B T}} dv_n \equiv \rho(v_n) dv_n. \qquad (B9)$$

The function $\rho(v_n)$ is simply the Maxwell velocity distribution as given in Eq. (B1). Furthermore, as long as the energy-momentum relation in Eq. (B3) holds, the results deduced above hold valid for NVT systems. Indeed, the results have been demonstrated in the dissolution dynamics of NaCl nanocrystal in liquid water [68], where strong particle-particle interactions present in a NVT system.

## Appendix C

In this appendix, we show the effects of perturbations/small changes of the energy barriers height on the calculated transmission coefficients of H atoms. Two sine-shaped energy barriers, one represents for the original barrier and the other for the perturbations are considered, with the analytic form as follows:

$$V_0(x) = U_0 \sin(\frac{\pi}{2}x) \quad (C1)$$

$$\delta V(x) = U_1 \sin(\pi x) \quad (C2)$$

where the parameters $U_0 = 0.2$ eV, $U_1 = 0.04$ eV. They are shown as below, with the perturbed barrier having the form:

$$V_1(x) = V_0(x) + \delta V(x) = U_0 \sin\left(\frac{\pi}{2}x\right) + U_1 \sin(\pi x) \quad (C3)$$

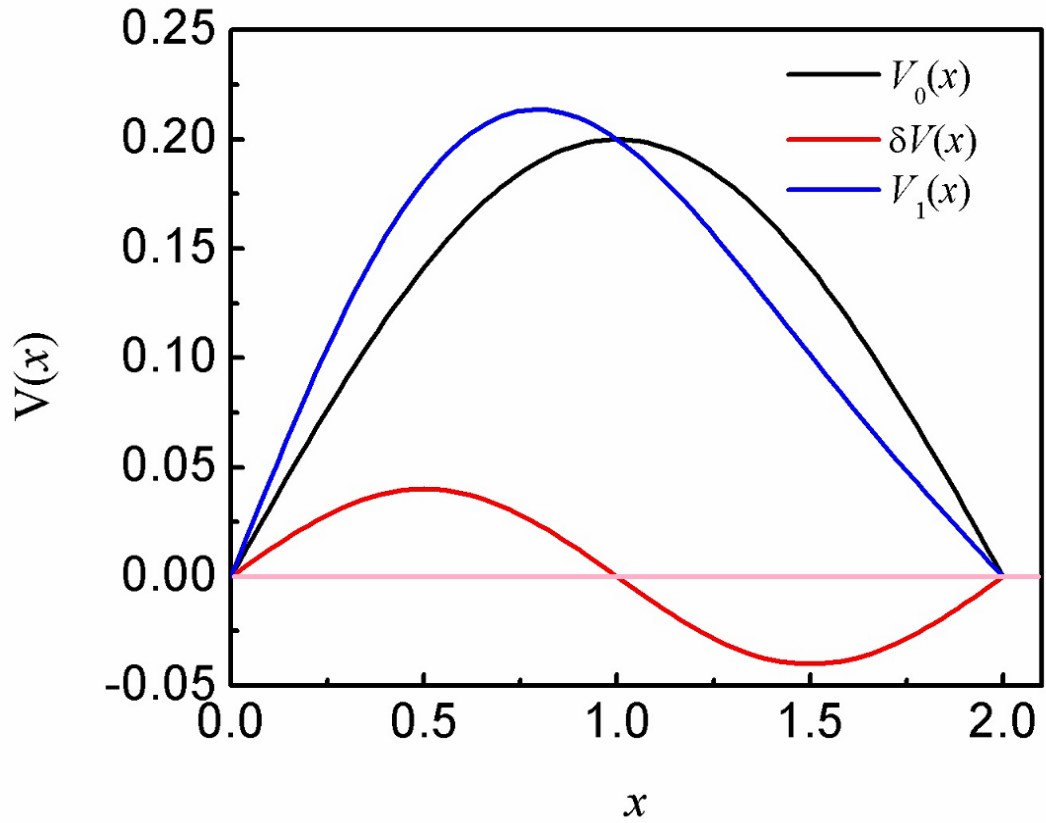

**Figure C1** Sine-shaped energy barriers considered for the evaluation of perturbations on transmission coefficient calculations.

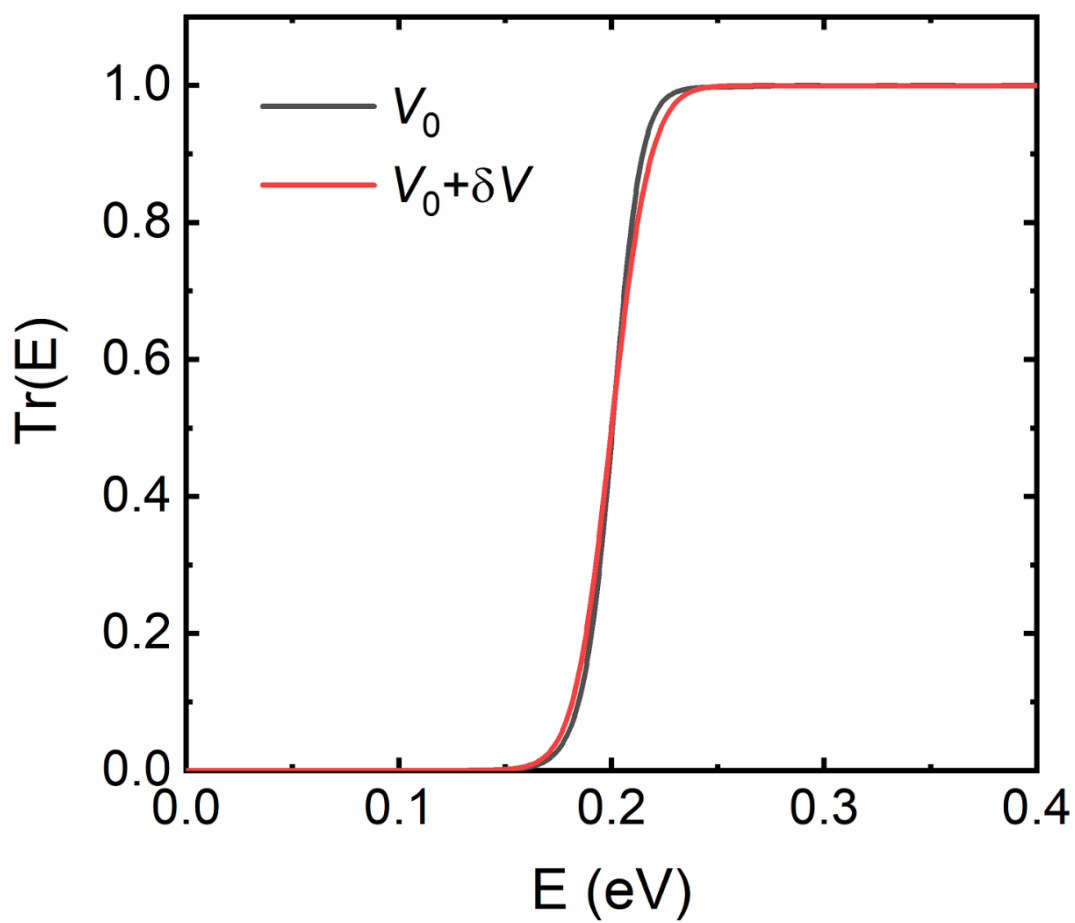

**Figure C2** Calculated transmission coefficient of H passing across the original ($V_0$) and perturbed ($V_0 + \delta V$) energy barriers.

The results calculated using the TM method are shown in Fig. C2. It is clear that perturbations/small changes to the original energy barrier would only result in minor modifications on the transmission coefficient.

**Table 1.** Calculated adsorption energies ($E_{ads}$), zero-point energies (ZPE), and the H-Pt bond lengths ($d_{PtH}$) of typical configurations of H adsorption on Pt(111). For each configuration, the PBE data of $E_{ads}$ without ZPE corrections are in parentheses.

|  | top | fcc | hcp |
|---|---|---|---|
| $E_{ads}$ (eV) | 2.596 (2.776) | 2.662 (2.801) | 2.620 (2.753) |
| ZPE (eV) | 0.180 | 0.139 | 0.133 |
| $d_{Pt-H}$ (Å) | 1.555 | 1.866 | 1.865 |

**Table 2.** Probabilities of barrier-crossing for H diffusion at 300 K, calculated by the transfer matrix (TM) and the WKB method, for different barriers $V(x)$. The total probability is divided into two parts contributed by two energy intervals, with $E = E_b$ being the border line. For H/Pt(111), the barrier heights $E_b$ were obtained by DFT-NEB method without the correction of zero-point energy.

| Barriers $V(x)$ | Transfer Matrix | | Semi-classical (WKB) | |
|---|---|---|---|---|
|  | $E < E_b$ | $E \geq E_b$ | $E < E_b$ | $E \geq E_b$ |
| rectangular ($E_b$ = 0.5 eV, b =3 Å) | $1.2938 \times 10^{-12}$ | $7.4105 \times 10^{-9}$ | $1.3442 \times 10^{-10}$ | $2.0307 \times 10^{-8}$ |
| top-fcc ($E_b$ = 0.11 eV) | 0.0148 | 0.0353 | 0.0180 | 0.0379 |
| top-hcp ($E_b$ = 0.13 eV) | 0.0084 | 0.0130 | 0.0094 | 0.0157 |
| top-hcp-hcp-top ($E_b$ = 0.13 eV) | 0.0043 | 0.0119 | 0.0039 | 0.0157 |

**Table 3.** Parameters $V(\eta)$ and $\lambda$ of Eq. (8) to describe the maxima of effective energy barriers ($E^*_{bm}$) displayed in Fig. 7.

|  | $V(\eta)$ (eV) | $\eta$ (Å) | $\lambda$ |
|---|---|---|---|
| $E_b^*$_Classical | $2.318 \times 10^{-2}$ | 0.239, 1.601 | 1.298 |
| $E_b^*$_ZPE Corrected | $2.489 \times 10^{-2}$ | 0.275, 1.467 | 2.138 |

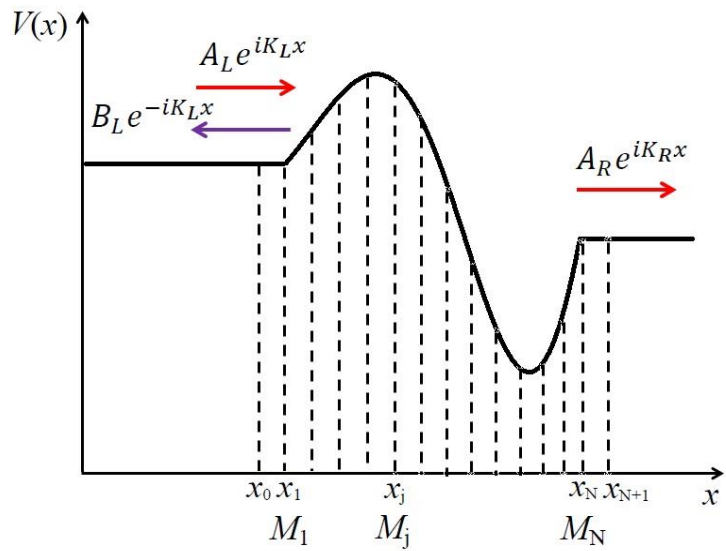

**Figure 1.** Schematic diagram for the calculation of particle transmission coefficient in a one-dimensional potential field based on the method of transfer matrix.

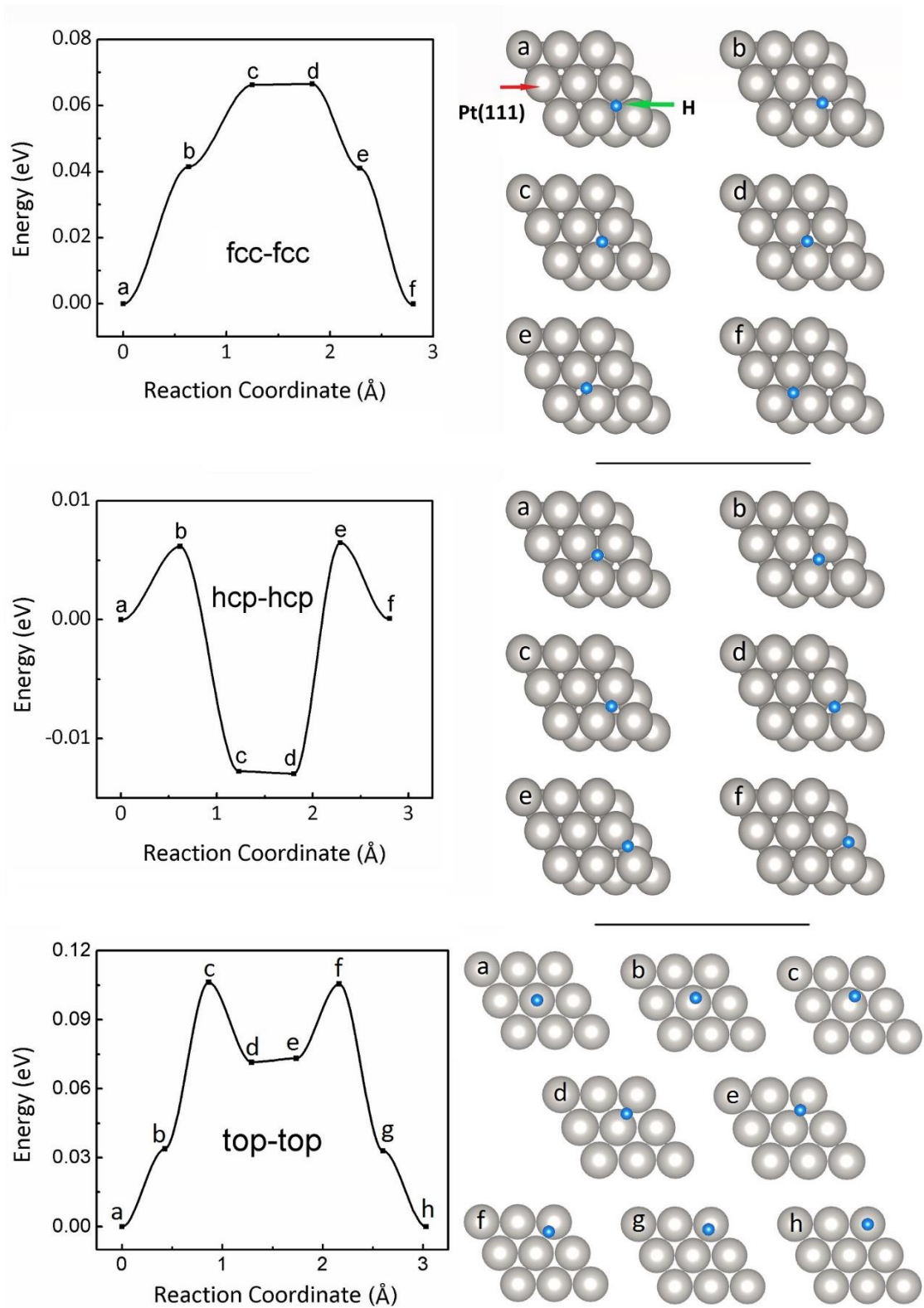

**Figure 2.** The optimal energy paths for H diffusion between the typical sites of Pt(111): fcc-fcc, hcp-hcp, top-top. The letters on the energy curves (left panels) have one-to-one correspondence to the atomic configurations on the right.

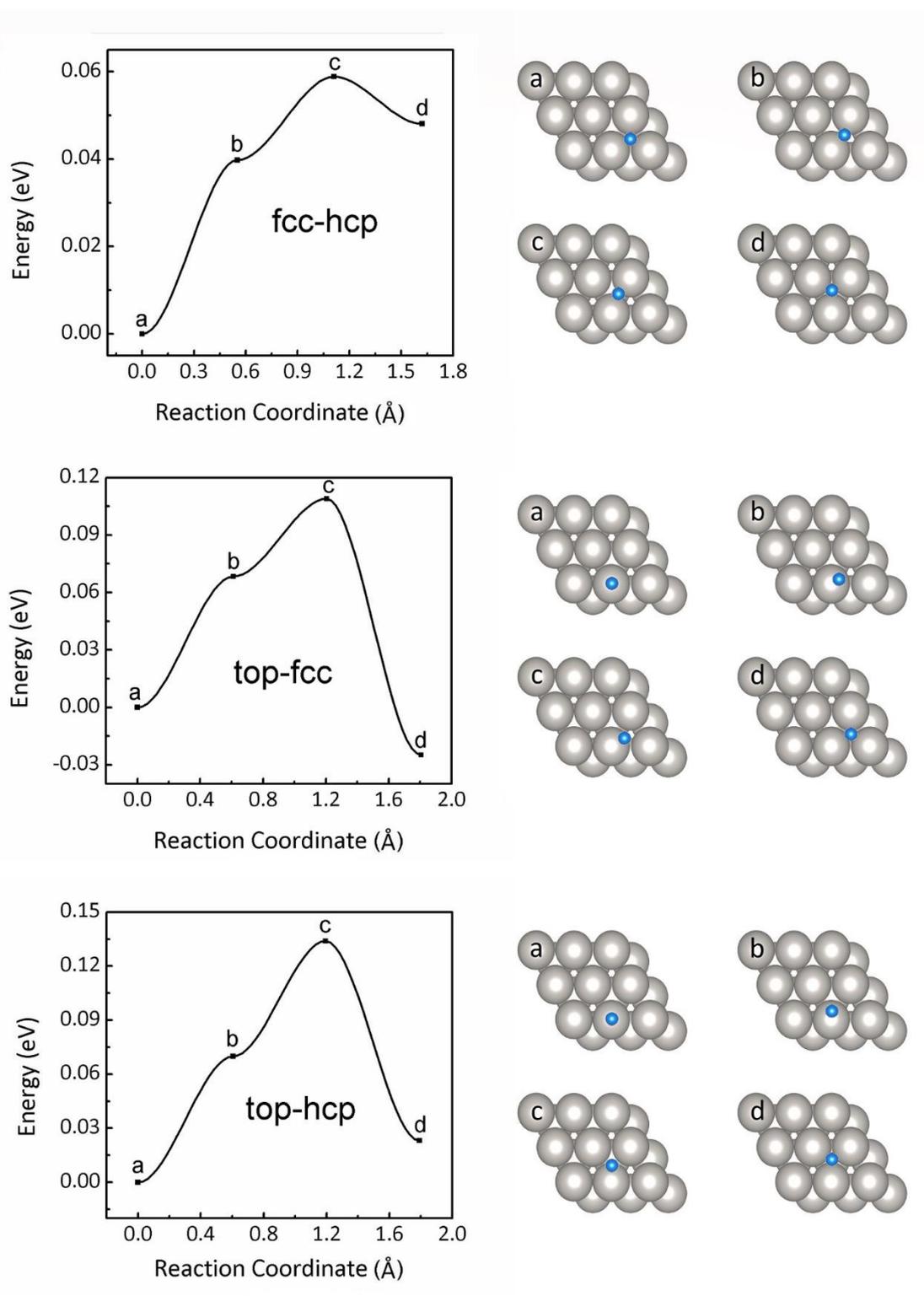

**Figure 3.** Similar to Fig. 2 but for the diffusion between different sites of Pt(111): fcc-hcp, top-fcc, and top-hcp.

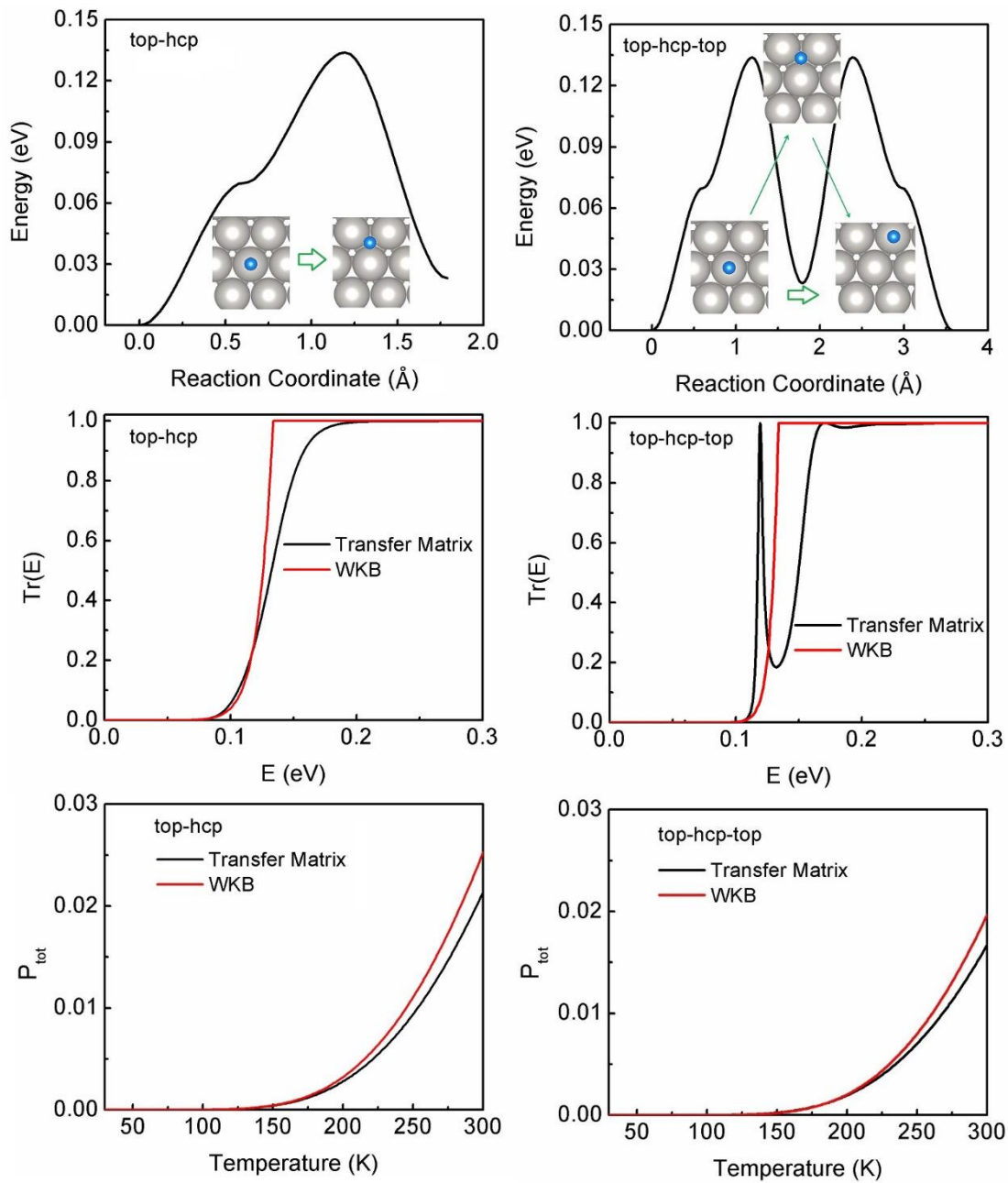

**Figure 4.** The potential energy pathways, the corresponding transmission coefficients Tr(E), and the total barrier-crossing probability ($P_{tot}$) as a function of temperature, for the diffusion paths top-hcp (left panels), and top-hcp-top (right panels). The results obtained by TM and WKB method are shown for comparison.

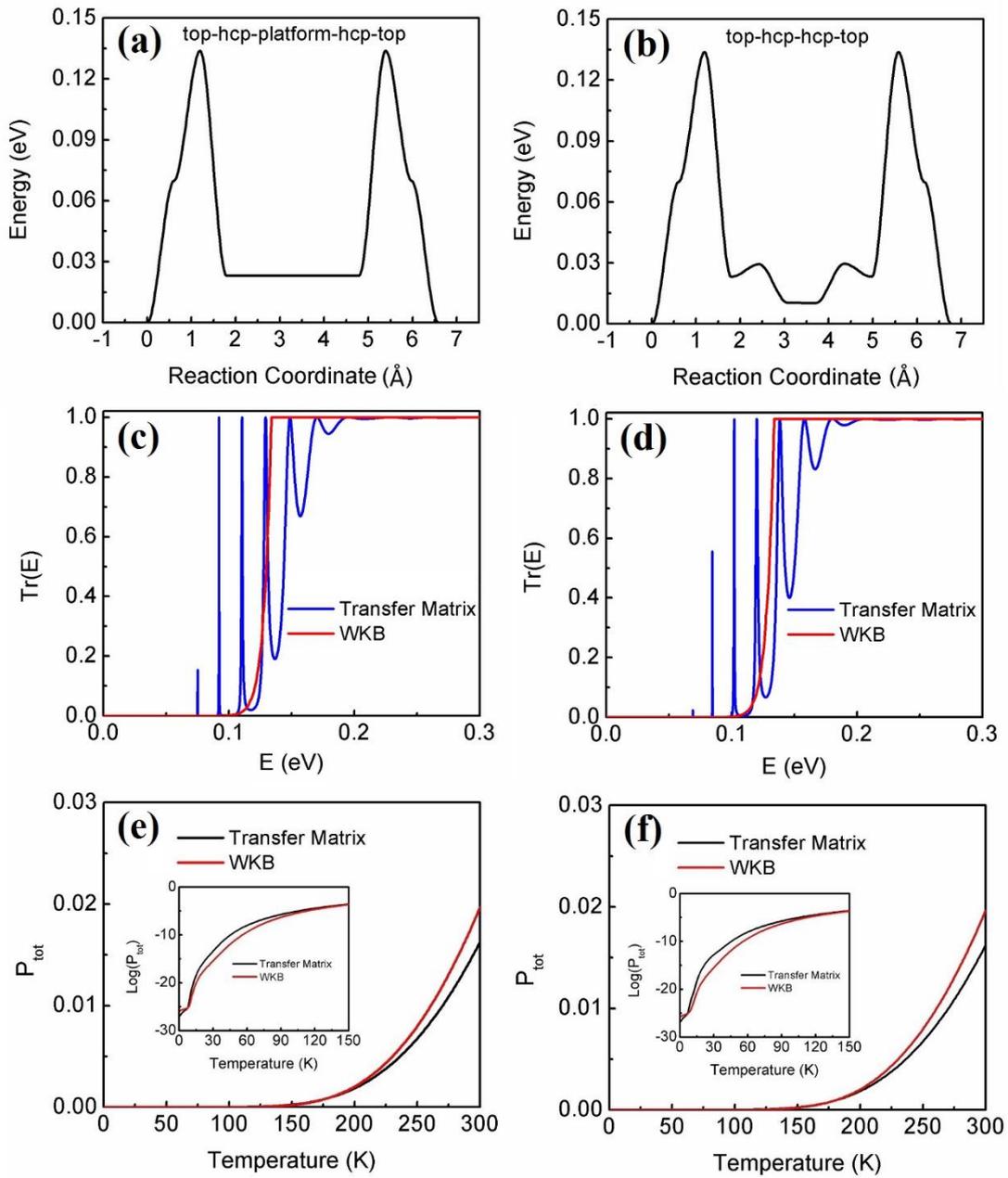

**Figure 5.** Transmission properties of H atoms along the constructed double barrier (top-hcp-platform-hcp-top, left panels) and the realistic approximate double barrier (top-hcp-hcp-top, right panels), calculated using the TM and WKB method.

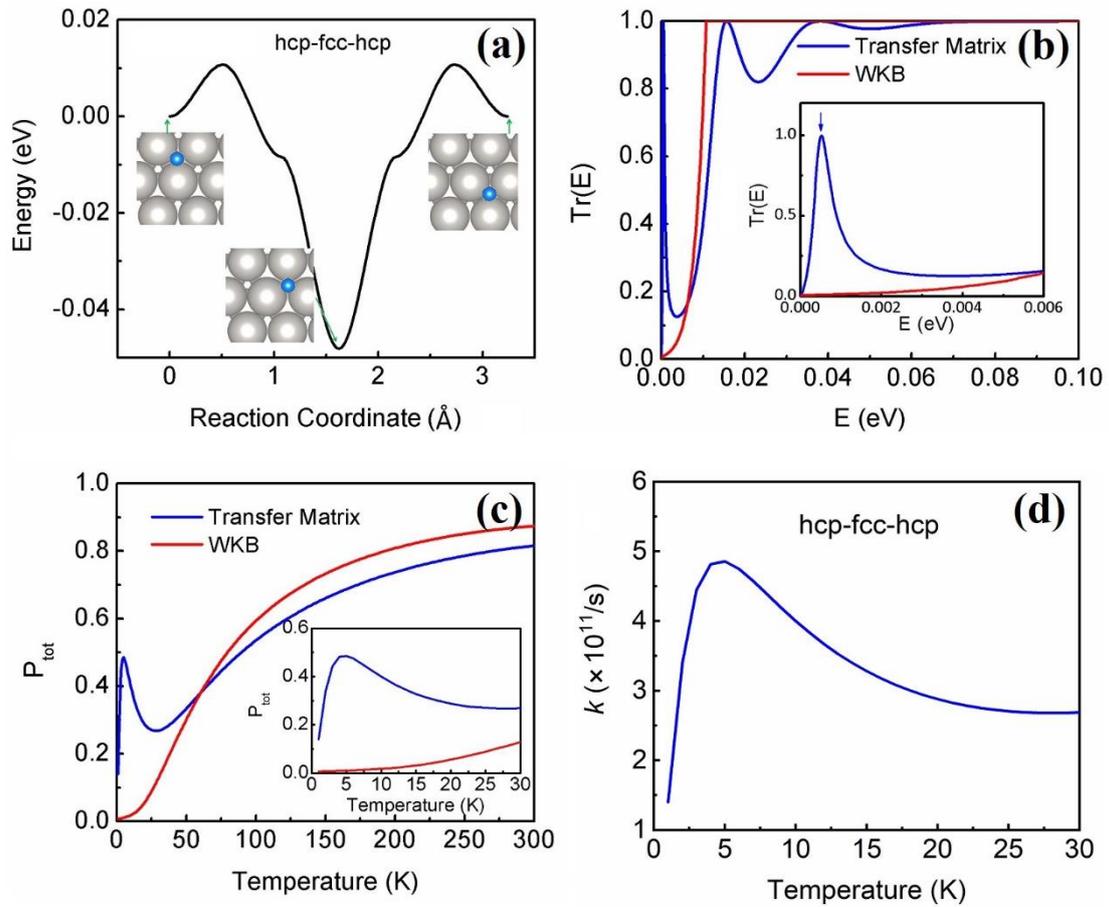

**Figure 6.** Panels (a-c): Similar to Fig. 4 but for the diffusion of H along the path hcp-fcc-hcp. (d): Temperature dependence of rate constant *k*.

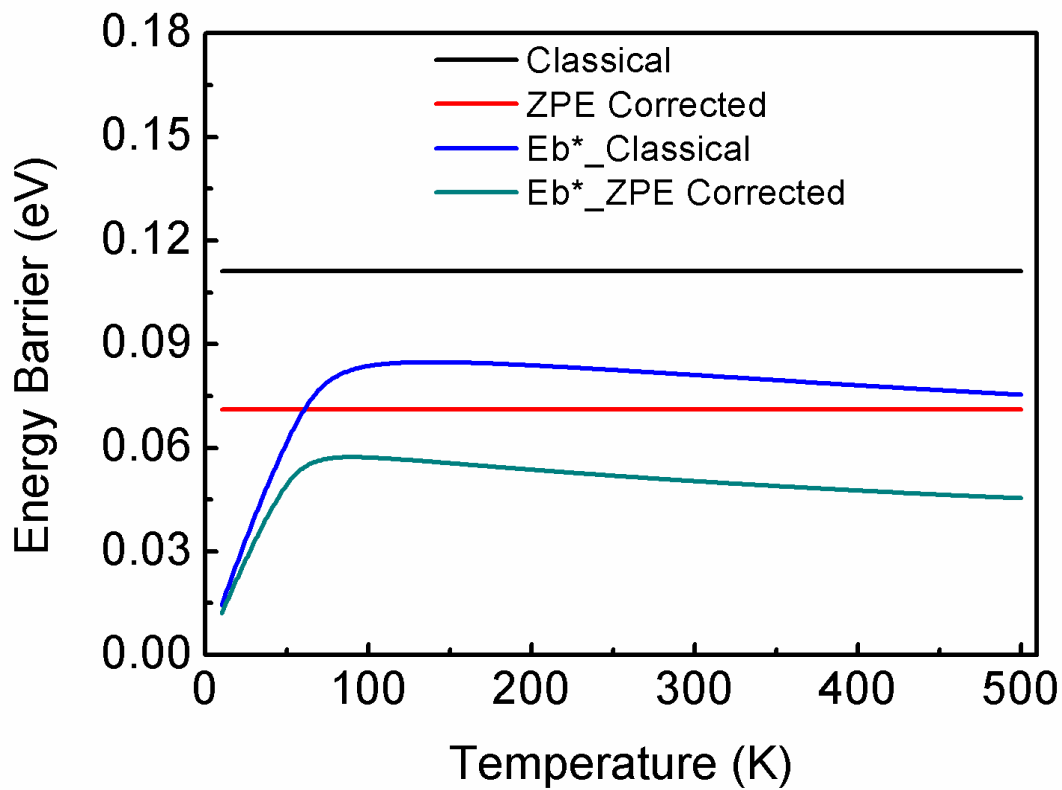

**Figure 7.** The energy barrier for the diffusion of H along the path top-fcc on Pt(111). The original barrier height calculated using adiabatic approximation is referred to as "Classical", and the data corrected by ZPE as "ZPE Corrected". The corresponding effective barriers ($E_b^*$) due to NQEs are presented as a function of temperature.